\documentstyle[preprint,aps,epsf]{revtex}
\tighten
\newcommand{\intsum}{\makebox[0cm][l]{$\, \sum$} \int}
\begin{document}

\draft

\title{
Nd elastic scattering as a tool to probe properties of 3N forces
}

\author{
H.~Wita\l{}a$^1$, 
W.~Gl\"ockle$^2$, 
J.~Golak$^{1,2}$, 
A.~Nogga$^2$, 
H.~Kamada$^2$\footnote{present address: Institut f\"ur Strahlen- und Kernphysik der Universit\"at  Bonn
Nussallee 14-16, D53115 Bonn, Germany}, 
R.~Skibi\'nski$^1$,
J.~Kuro\'s-\.Zo{\l}nierczuk$^1$.
}
\address{$^1$Institute of Physics, Jagellonian University,
                    PL-30059 Cracow, Poland}
\address{$^2$Institut f\"ur Theoretische Physik II,
         Ruhr Universit\"at Bochum, D-44780 Bochum, Germany}

\date{\today}
\maketitle

\begin{abstract}
Faddeev equations for elastic Nd scattering have been solved using modern
NN forces combined with the Tucson-Melbourne two-pion exchange three-nucleon force, 
with  a modification
thereof closer to chiral symmetry and the Urbana IX three-nucleon force. Theoretical
predictions for the differential cross section and several spin observables
using NN forces
only and NN forces combined with three-nucleon force models are compared to each other and
to the existing data. A wide range of energies from 3 to 200 MeV is covered. Especially at
the higher energies striking three-nucleon force effects are found, 
some of which are
supported by the still rare  set  of data, some are in conflict with data  and thus
very likely point to defects in those three-nucleon force models.
\end{abstract}

\narrowtext

\section{Introduction}
\label{secIN}

One major goal in nuclear physics is to establish the properties of nuclear forces and
to understand nuclear phenomena by solving the many-
nucleon Schr\"odinger equation driven by those elementary nuclear forces. Meson theory
had an important impact for the construction
of nuclear forces (both NN and 3N forces), but it lacks systematics like it would be given
by an expansion parameter and the meson-nucleon
vertices had to be parametrised in an ad-hoc manner. Nevertheless  the one-pion exchange
is undisputed and the most advanced formulation of
meson-exchanges in the so called full Bonn potential~\cite{ref1}
 is remarkably successful.
Because of its energy dependence - a consequence of deriving
it by old fashioned time ordered perturbation theory - it is not useful in $ A > 2$ systems.
 Energy independent one-boson exchange versions
thereof, however, are useful~\cite{ref2,ref3}. 
In addition more phenomenological NN potentials have been
 constructed with the aim to describe the rich set
of experimental NN data as precisely as possible. This leads to an often called new generation
of realistic NN potentials: AV18~\cite{AV18}, 
CD Bonn~\cite{CDBONN}, Nijm~I, II and 93~\cite{ref3}. 
They describe the NN data set with an unprecedented precision
of $\chi^2$ per data point very close to one. Very
recently an updated CD Bonn~\cite{ref4} appeared, which takes newest data 
into account but has not been used in the present article. 
An upcoming approach
to construct nuclear forces in a systematic manner is chiral perturbation
theory~\cite{ref5,ref6,ref7,ref8,ref9,ref10,ref11}. First applications 
to three- and
four-nucleon systems have been done~\cite{ref12}. 

In recent years 
it became possible 
to solve exactly three- and four-nucleon bound states using standard
integration and differentiation methods~\cite{ref13,ref14}. 
Stochastic techniques allow to go beyond A=4
 and right now low energy states of nuclei up to A=8 are
under control~\cite{ref15,Wiringanew}. In all cases those realistic NN forces 
fail to provide the experimental
 binding energies; there is clear cut underbinding.
For instance this amounts to 0.5-1 MeV in case of three nucleons and to 2-4 MeV in case of $^4$He.
 A natural further step is the consideration of 3N
forces. This is an even harder theoretical challenge and presently the most often used
 dynamical process is the $\pi-\pi$ exchange between three
nucleons with an intermediate excited nucleon state, the $\Delta$~\cite{ref16}.
This is augmented by
 further ingredients of various types as will be detailed
below. By properly adjusting parameters one can achieve correct 3N and 4N binding energies
 and reaches even a fairly successful description of
low energy bound states energies of up to A= 8~\cite{ref16a}. 
However, in the latter case the results point to an insufficient 
spin-orbit splitting of nuclear levels in light nuclei as e.g. 
in $^5$He~\cite{ref15}. 
This may be caused by a wrong spin structure 
of present day 3NF's or by not well enough established $^3P_j$ NN force components.  It will be interesting to see in the future the predictions
of nuclear forces based on chiral perturbation theory.

Though this first signal on 3N force effects resulting from discrete states is important,
 a more detailed investigation of 3NF properties can
be carried through  in scattering processes, where a rich set of spin observables is available.
The tremendous advance in computational resources allowed in recent years to make exact
 predictions  for three- nucleon scattering  using
nuclear forces in all their complexities~\cite{ref17}. Also experimentally 
one can access nowadays
 spin observables in Nd scattering where in the
initial states the deuteron and / or  the nucleon is polarized and after the reaction
 also the polarization of the outgoing particles can
be  measured~\cite{ref18,ref18a,ref18b,ref18c,ref18d,ref18e,ref18f,ref28a}. 
This leads to a very rich spectrum of observables in Nd elastic scattering
and the Nd break-up processes. Such a set of spin
observables will be a solid basis to test the 3N Hamiltonian. Using available model
 Hamiltonians one can provide guidance in selecting specific
observables and energies which are most appropriate to see 3NF properties. It is the aim
 of this article to do exactly that and to compare
the theoretical predictions with already existing data.

In section~\ref{secII} we review briefly our 3N scattering formalism and display 
the 3NF model forces
 which are presently en vogue and which we use.
Some technical details referring to the partial wave decomposition of the momentum space
 representation of the Urbana~IX 3NF~\cite{ref19} are given in the
Appendix. In this article we restrict ourselves to elastic Nd scattering and refer to a
 forthcoming article for the break-up process. Our
predictions for various nuclear force combinations and the comparison to available data are
 given in section~\ref{secIII}. We conclude in section~\ref{secIV}.

\section{Scattering Formalism and 3NF Models}
\label{secII}

We refer to~\cite{ref17} for a general overview on 3N scattering and specifically 
our way to formulate
it. For the inclusion of 3NF´s we found meanwhile a more efficient 
way~\cite{ref21}. 
It is a direct
generalization of what is being used for the 3N bound state~\cite{ref22}. 
We define an amplitude $T$ via our
central equation
\begin{equation}
\label{eqT}
T \ = \ t \, P \, \phi \ 
+ \ ( 1 + t G_0 ) \, V_4^{(1)} \, ( 1 + P ) \, \phi \
+ \   \ t \, P \, G_0 \, T \
+ \ ( 1 + t G_0 ) \, V_4^{(1)} \, ( 1 + P ) \, G_0 \, T
\end{equation}
The initial channel state $\phi$ occurring in the driving terms is composed of a deuteron and a
momentum eigenstate of the projectile nucleon. The NN $t$-operator is denoted by $t$, 
the free 3N propagator by $G_0$ and $P$ is the sum of a cyclical and an anticyclical 
permutation of three particles. The 3N force $V_4$ can always be decomposed into 
a sum of three parts

\begin{equation}
\label{eqV4}
V_4 = V_4^{(1)} + V_4^{(2)} + V_4^{(3)} ,
\end{equation}
where $V_4^{(i)}$ is symmetrical under the exchange of the nucleons $j k$ 
with $j \ne  i \ne k$.
As seen in Eq.~(\ref{eqT}) only 
one of the three parts occurs explicitely, the others via the permutations
contained in $P$. 
The physical break-up amplitude is given via
\begin{equation}
U_0 = ( 1+ P) T .
\label{eqU0}
\end{equation}

 The Faddeev-like integral equation~(\ref{eqT}) has the nice property 
that its iteration inserted into Eq.~(\ref{eqU0}) yields immediately
the multiple scattering series, 
which gives a transparent insight into the reaction
mechanism. Here in this article we concentrate on elastic scattering, 
whose amplitude is given by
\begin{equation}
U \ = \  P G_0^{-1} \ + \ P T \ + \ V_4^{(1)} \, ( 1 + P ) \, \phi \ + \
V_4^{(1)} \, ( 1 + P ) \, G_0 \, T
\label{eqU}
\end{equation}
The first term is the well known single particle exchange diagram,
then there are terms where either $V_4$ or the $t$'s interact once and then the remaining parts
result from rescattering among the three particles. Again inserting the iteration 
of $T$ as given in Eq.~(\ref{eqT}) into Eq.~(\ref{eqU}) 
yields a transparent insight~\cite{ref23}.

 The definition of the various spin observables can be found 
in~\cite{ref17,ref24}. They
have the general form
\begin{equation}
\label{eqSPIN}
\langle S^\mu \rangle_f I = \frac16 \sum\limits_\nu \langle S^\nu \rangle_i 
{\rm Tr} \left( M S^\nu M^\dagger  S^\mu \right) ,
\end{equation}
where $I$ is the elastic cross section summed over the spin states in the final state, 
$M$ is the
physical elastic scattering amplitude related directly to $U$ and $ S^\mu$  is  
a suitable set of 3N spin operators. 

We shall encounter nucleon and deuteron vector analyzing powers $ A_y (N)$   
and $A_y (d)$  (i$T_{11}$),  where in 
the initial state either the nucleon (N) or the deuteron (d) is 
vector polarized. 
Further, the deuteron can
be tensor polarized in the initial state leading to the three tensor analyzing powers $T_{2k}$
(k = 0,1,2). Also both particles can be polarized in the initial state leading to very many spin
correlation coefficients $C_{\alpha, \beta}$, where $\alpha$ refers to the spin directions
 of the nucleon and beta 
  to vector and
tensor polarizations of the deuteron. Further information on the dynamics can be found in spin
transfer coefficients $ K_\alpha^{\beta '}$, where $\alpha$ describes either a polarized nucleon 
or  a polarized deuteron in the initial state 
and  $\beta '$ similarly the polarization for a particle in the final state. 
Of course all those quantities depend on the scattering angle.

Our nuclear model interaction consists of one of the NN forces mentioned 
in the introduction and a 3NF. For the 3NF we use the $2 \pi$-exchange 
Tucson Melbourne (TM) model,
a modified version thereof and the Urbana~IX force.
The TM model~\cite{ref25}
has been around for quite some time. 
It is based on a low momentum expansion of the $\pi-N$ off (the- mass) 
shell scattering amplitude.
It has the form~\cite{ref25}
\[
V_4^{(1)} = { 1 \over {( 2 \pi )^6} } \, 
            { g^2_{\pi N N} \over { 4 m_N^2} } \,
          {  { {\vec \sigma}_2 \cdot {\vec Q} } \over {\vec{Q}^{\, 2} + m_\pi^2} } \,
          {  { {\vec \sigma}_3 \cdot {\vec {Q '}} } \over {\vec{Q '}^{\, 2} + m_\pi^2} } \,
          H \left( \vec{Q}^{\, 2} \right) \, H \left( \vec{Q '}^{\, 2}  \right) 
\]
\begin{equation}
\left\{ \vec{\tau}_2 \cdot \vec{\tau}_3 \left(  a \,   + \,  b \vec{Q} \cdot \vec{Q '} \,
                                                + \, c ( \vec{Q}^{\, 2} +  \vec{Q '}^{\, 2} ) \right)
               \ + \ d \, i \, \vec{\tau}_3 \times \vec{\tau}_2 \cdot  \vec{\tau}_1 \,
                  \vec{\sigma}_1 \cdot {\vec Q} \times \vec{Q '}  \right\} .
\label{eqTM3NF}
\end{equation}

The elements of the underlying Feynman diagram are obvious: the two pion propagators depending on
the pion momenta $\vec{Q}$  and $\vec{Q '}$,  
the two $\pi N N$ vertex amplitudes and most importantly the parametrisation
of the $\pi N $ amplitude inside the curly bracket which is combined 
with the isospins  $\vec{\tau}_2$  and $\vec{\tau}_3$ of
the two accompanying nucleons. On top of all that there is a strong form factor parametrisation
given by

\begin{equation}
 H \left( \vec{Q}^{\, 2} \right) \, = \,
\left(  { {\Lambda^2 - m_\pi^2} \over {\Lambda^2 + \vec{Q}^{\, 2}} }  \right)^2
\label{eqH}
\end{equation}
In what we denote by the TM 3NF we use the original parameters
$ a = 1.13 / m_\pi $, 
$ b = -2.58 / m_\pi^3 $, 
$ c = 1.0 / m_\pi^3 $, 
$ d = -0.753 / m_\pi^3 $.
They incorporate among others 
the physics resulting from an intermediate  $\Delta$ in a static 
approximation. The cut-off parameter $\Lambda$ 
is used to adjust the $^3H$ binding 
energy separately for different NN forces~\cite{ref26}.
For the convenience of the reader we show the $\Lambda$--values 
in Table~1. 
Of course in a meson exchange picture additional processes should be added containing 
other meson exchanges like $\pi - \rho$,
$\rho - \rho$; also different intermediate excited states might play a role. To some extent
3NF models with respect to those extensions have already been developed 
and applied~\cite{ref27,ref28,ref29,ref30}. 
Further studies should be performed.

The parametrisation of the TM 3NF has been criticized somewhat, since it violates chiral
symmetry~\cite{ref31,ref32}. A form consistent with chiral symmetry (though not a complete 
one to that order in the appropriate power counting) is obtained
by modifying the $c$-term so that the long-range part is absorbed into the $a$-term,
leading to a new $ a' \equiv a - 2m_{\pi}^2 c = -0.87 / m_\pi $~\cite{ref31,ref32} 
what essentially means 
a change of sign for $a$ and that the short range part is dropped. This form will
be called TM' later on. The corresponding $\Lambda$-value 
when used with the CD Bonn potential
 is $\Lambda = 4.593 \, m_\pi$.

The two-meson exchange 3NF has also been studied 
by Robilotta {\em et al.}~\cite{ref33}
leading to the Brazilian
3NF. It is similar to the one of TM and also the results gained 
for low energy Nd elastic scattering observables~\cite{ref34}
are similar to the ones for the TM 3NF. In this article we do not 
take that force into account.
Instead we included the Urbana IX 3NF~\cite{ref19}, 
which is heavily used in the Urbana-Argonne
collaboration. It will be interesting to see its effects for 3N scattering observables. At very
low energies it has been used in that context before
by the Pisa group~\cite{ref35}. 
That force is based on the old Fujita-Mijazawa ansatz~\cite{ref36} 
of an intermediate 
$\Delta$ occurring in the two-pion exchange and
augmented by a spin independent short range piece. It has the form
\[
V_4^{(1)} = A_{2 \pi} \, \left[ \{ X_{12} ,  X_{13} \} \,
               \{ \vec{\tau}_1 \cdot \vec{\tau}_2 , \vec{\tau}_1 \cdot \vec{\tau}_3 \} \,
 + \, \frac14 \, [ X_{12} ,  X_{13} ] \, 
    [ \vec{\tau}_1 \cdot \vec{\tau}_2 , \vec{\tau}_1 \cdot \vec{\tau}_3 ] \, \right]  \ 
\]
\begin{equation}
+ \ U_0 \, T_\pi^2 (r_{12}) \, T_\pi^2 (r_{13}) , 
\label{eqURBANA3NF}
\end{equation}
where 
\begin{equation}
 X_{i j} = Y_\pi (r_{i j}) \,
\vec{\sigma}_i \cdot \vec{\sigma}_j \ + \ T_\pi (r_{i j}) \, S_{i j} ,
\label{eqXIJ}
\end{equation}
with 
\begin{equation}
Y_\pi (r) \, = \, 
{ e^{- m_\pi r} \over {  m_\pi r } } \, \left( 1 - e^{- c r^2} \right) ,
\label{eqYPI}
\end{equation}
and
\begin{equation}
T_\pi (r) \, = \, \left[ 1 + { 3 \over {m_\pi r} } + { 3 \over {(m_\pi r)^2} } \right] \,
{ e^{- m_\pi r} \over {  m_\pi r } } \, \left( 1 - e^{- c r^2} \right)^2 ,
\label{eqTPI}
\end{equation}
and where 
\begin{equation}
S_{i j} \, = \, 3 \,  \vec{\sigma}_i \cdot \hat{r}_{ij} \, 
                      \vec{\sigma}_j \cdot \hat{r}_{ij} \ - \
\vec{\sigma}_i \cdot \vec{\sigma}_j
\label{eqS}
\end{equation}
is the tensor operator.

Since we work in momentum space and in a partial wave expansion the form 
given in Eq.~(\ref{eqURBANA3NF}) has to be
rewritten. We could follow the steps laid our before for the corresponding representation of the
TM 3NF~\cite{ref37} and delegate all that to the Appendix. 

Since there is no apparent consistency of the mostly phenomenological 
realistic NN forces and the 3NF models we test various combinations thereof.
In all cases, however, we require that the particular 
choice for the 2N interaction and the 3NF should reproduce the experimental triton binding 
energy. Some of the 3N observables scale with the triton binding energy~\cite{Witalascaling}. 
The adjustment to the triton binding energy has the advantage that our investigation
is not misled by these scaling effects.

With respect to the intermediate $\Delta$ one should say that very likely the static approximation
is not justified and the $\Delta$ should be allowed to propagate like the nucleon. This has been
pursued intensively for instance by the Hannover group~\cite{ref38} 
and recent work has 
been also devoted to the 3N continuum~\cite{ref39}. 

In view of all that it is quite 
clear that our present study is not at all complete but can at most provide some insight, 
what kind of effects specific 3NF models might generate.
As we shall see effects of that sort are needed, since NN force only predictions often
fail to describe the data, especially at the higher energies. These challenges call for a
systematic approach and at least for the leading spin structures chiral perturbation theory
might be a good candidate~\cite{ref6}. This is left to a future investigation. Here we 
concentrate on the current models and show their strengths and failures.

\section{Predictions of 3NF Effects and Comparison to Data}
\label{secIII}

Since we would like to cover a wide range 
of incoming neutron energies from below the nd breakup threshold up 
to 200 MeV it is necessary to take a sufficient number of 
partial wave states into account in order to get converged 
solutions of the Faddeev equations. 
In all calculations presented in this paper we went up to
the two-nucleon subsystem total angular momentum 
$j_{max}=5$. This corresponds to a maximal number of $142$ partial wave states 
(often called channels) in the 3N system. We checked that the convergence 
has been achieved by looking at the
results obtained for $j_{max}=6$, which increases the number 
of channels to 194. This convergence check refers to a calculation
without a 3NF. The inclusion of 3NF's has been carried through
for all total angular momenta of the 3N system up to $J=13/2$. 
These high angular momenta are required at the higher energies $\ge 100$ MeV. 
The longer ranged 2N interactions require states up to  $J=25/2$
at the higher energies
in order to get converged results.

A phenomenological criterium for 3NF effects is that the data lie outside the spread of NN
force predictions only. In the following figures we shall always include a shaded band
(called ``band1''),
which covers the predictions of the AV18, CD-Bonn, NijmI, II and 93 NN forces. Unfortunately
we cannot include the pp-Coulomb force 
in our approach and thus have to live with some theoretical
uncertainty when comparing to pd-data. At the higher energies, however, those effects should
be small. Also in case of AV18 we do not take the various electromagnetic corrections into
account, which leads for example to a slightly wrong deuteron binding energy ($E_d$ = 2.242
instead of 2.225 MeV).
This, however, has only a small effect on our results, which is mostly of 
kinematical origin, since the phase-shifts obtained without those 
additional terms differ only slightly from the standard ones.
The kinematical effects are seen predominantly in the breakup process, 
where this small defect in the deuteron 
binding energy leads to correspondingly small shifts in peak structures
like for instance final state interaction peaks.

We shall combine another group of curves into a band.
The TM 3NF can be combined with the five NN forces. In all cases the cut-off value 
$\Lambda$ in Eq.~(\ref{eqH}) has been adjusted separately for
each NN force to the $^3$H binding energy~\cite{ref26}. Since that interplay 
is a purely phenomenological step the outcome is 
theoretically not under control 
and we combine all the results 
into a second band (called ``band2'' in the following). 
Next we want to compare the TM 3NF and the
modified TM', which is more consistent 
with chiral symmetry. We combine 
it with CD-Bonn and show the
CD-Bonn+TM' prediction as a dashed curve.
 Finally we compare
the TM and the Urbana IX 3NF's and combine 
them with AV18. The combination AV18+URBANA~IX will appear as a solid
curve.
There are clearly 
more combinations possible but it is 
sufficient to get an orientation on the magnitudes of expected effects.

\begin{figure}[h!]
\leftline{\mbox{\epsfysize=185mm \epsffile{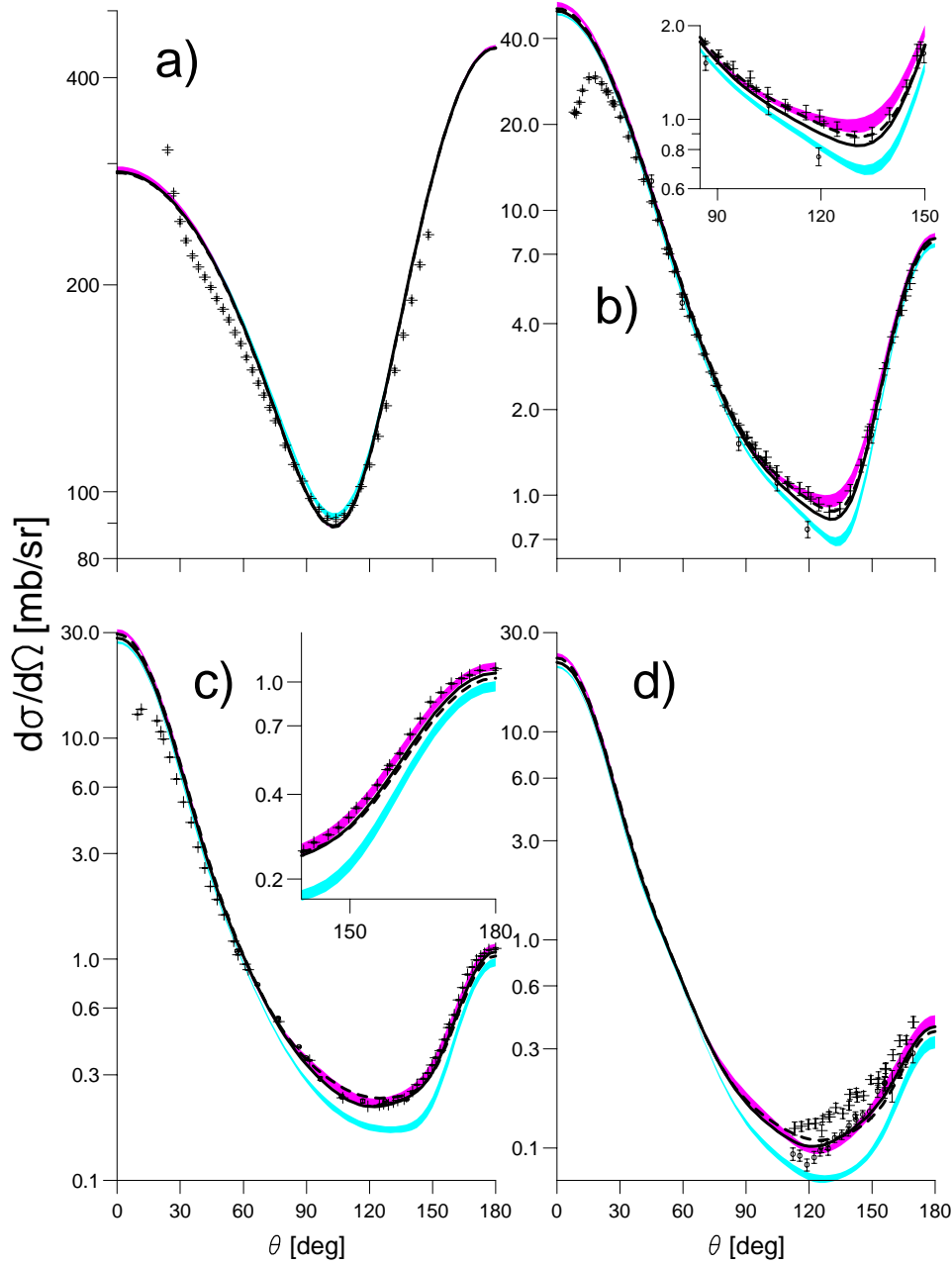}}}
\caption[ ]{The differential cross section in elastic Nd scattering 
at 3 (a), 65 (b), 135 (c) and 190 MeV (d).
Two bands are shown in each subfigure, the light shaded one contains
NN force predictions, the darker shaded one the NN force predictions+TM 3NF.
The solid curves are the AV18+URBANA~IX predictions. 
The dashed curves are the CD Bonn+TM' predictions.
Data at 3 MeV from~\protect\cite{ref41} (pd), at 65 MeV 
from~\protect\cite{ref42} (pd-crosses) and ~\protect\cite{ref51a} 
(nd-circles), 
 at 135 MeV from~\protect\cite{ref18} (pd-crosses),
~\protect\cite{sakamo} (pd-circles),  
 and at 190 MeV 
 from~\protect\cite{ref55a} (pd: crosses 181 MeV, circles 216.5 MeV).
In some cases error bars are not visible on the scale of the figure.}
\label{fig1}
\end{figure}

We begin with the differential cross section in Fig.~\ref{fig1}. 
The various NN force predictions
are rather close together with a small spread in the minima. Including the TM 3NF there is
again a small spread in the minima 
(practically negligible at 3 MeV) but the 
minima are shifted upwards,
rather well into the data~\cite{ref43} 
except at 3 MeV, where the 3NF prediction
is shifted slightly downwards. 
The phenomenon of shrinkage 
of the spread between different NN potential predictions
by including a 3NF is often called
a scaling phenomenon. It occurs at low energies and is related to 
the three-nucleon
binding energy, which by construction is common to all those curves in band2. 
The TM and TM' 3NF's together with CD-Bonn 
give slightly different predictions in the 
minima especially at the two highest energies. 
(The CD Bonn+TM prediction lying inside band2 is not shown.)
In the backward angular region they differ significantly and the 135 MeV 
precise
backward angular distributions data prefer the TM 3NF. (See insertion 
in Fig. 1; again the CD Bonn+TM prediction is not explicitely shown.)
On the other hand TM and URBANA~IX together with AV18 are very similar at 
the two higher 
energies  but differ significantly at $E$= 65 MeV. Certainly 
Coulomb force effects
should be taken into account at this energy before a final 
conclusion can be drawn.  The few nd data near the minimum 
would strongly disagree with all our 3NF predictions and a confirmation
(or rejection) would be highly desirable.
However, independent from  possible Coulomb force contributions, it is clearly 
seen, that even such a simple observable as the elastic scattering 
differential cross section exhibits large 3NF effects at higher energies. 
These effects are not trivial and depend not only on the 
incoming energy and the angle but also on the particular 3NF used. 
This calls for precise data for this observable in order to study the 3NF properties.

\begin{figure}[h!]
\leftline{\mbox{\epsfysize=185mm \epsffile{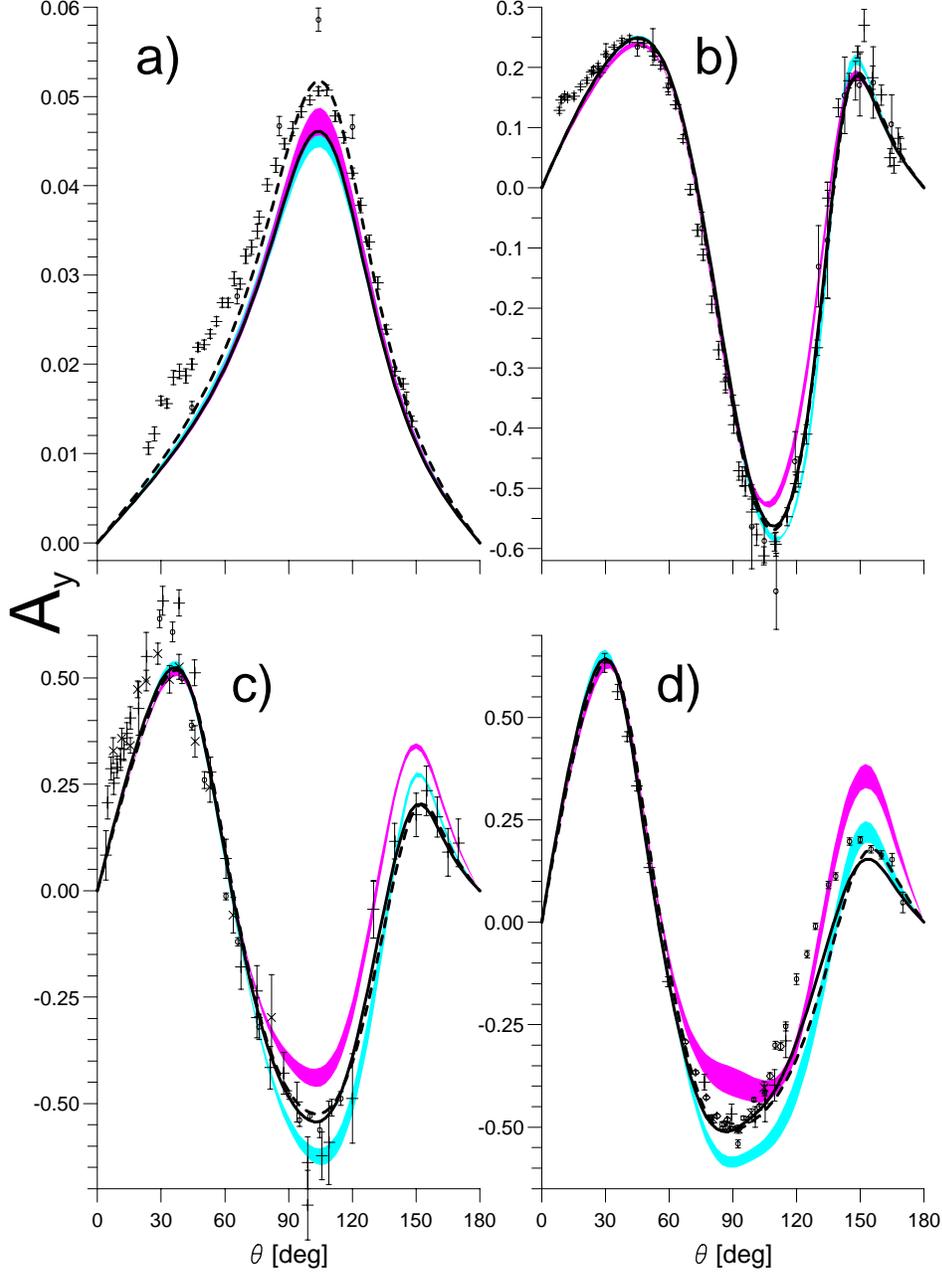}}}
\caption[ ]{The analyzing power $A_y (N)$ for elastic Nd scattering.
Curves and the sequence of energies as in Fig.~\protect\ref{fig1}. 
Data at 3 MeV from~\protect\cite{ref42} (pd-crosses) and 
~\protect\cite{ref44} (nd-circles), 
at 65 MeV from~\protect\cite{ref42} (pd-crosses) and 
~\protect\cite{ref51a} (nd-circles),
at 135 MeV from~\protect\cite{ref18d} (pd-circles 150 MeV), 
~\protect\cite{ref62} (pd-crosses 146 MeV),
~\protect\cite{ref63} (pd-x's 155 MeV),
and at 190 MeV from~\protect\cite{ref18d} (pd-crosses 190 MeV), 
~\protect\cite{ref60} (pd-circles 198 MeV),
~\protect\cite{ref61} (pd-squares 197 MeV).}
\label{fig2}
\end{figure}

Let us now regard a selection out of 
the many spin observables in elastic Nd scattering.
Fig.\ref{fig2} shows $A_y (N)$. 
The band for NN force predictions is always rather narrow,
whereas band2 for the lowest and highest energy 
is distinctly broader. The two bands
are separated predicting clearly 3NF effects 
especially at higher energies.
The TM and TM' predictions are distinctly 
different as well as the TM and Urbana~IX
predictions. It is interesting to note that here TM' with CD Bonn
and Urbana IX with AV18 are very similar (except at the lowest energy)
and predict
only small 3NF effects at 65 MeV, which are compatible with the $A_y$ data 
at this energy. 
At higher energies their effects become quite different 
from the TM ones. While in the region of the $A_y$ minimum around 
$\theta_{cm} \approx 110^o$ they increase $A_y$ as compared to the pure 
2N force predictions, their action decreases $A_y$ in the backward angular region, 
contrary to the action of the TM 3NF model, bringing the theory 
closer to the data. In this way the TM' and Urbana IX 3NF's seem to solve partially
the $A_y$ problem found at higher energies in ~\cite{ref18d,ref46}. 
At 3 MeV the clear discrepancy  of all theoretical 
predictions to the nd data of ~\cite{ref44} is seen. At such a low 
energy it is well known that Coulomb force effects are large for $A_y$ 
decreasing significantly its maximum 
when compared to nd data~\cite{ref49}. Thus also pd data lie very clearly 
(due to much smaller error bars) above all theoretical 
predictions. We see that this very well known low energy $A_y$ 
puzzle~\cite{ref50} cannot be solved by the 3NF models we are using 
in this article. A slightly 
increased maximum of $A_y$ for TM' is far too small to play 
any significant role and possibly the solution should be also 
sought in 
an improvement of the $^3P_j$ NN force components~\cite{ref51} to which low energy 
$A_y$ is very sensitive. We would also like to point to
the very recent result based on chiral perturbation theory
~\cite{ref12}, where $A_y$ can be described quite well 
in next-to-leading order (NLO). 
In that order of the power counting 3NF's do not yet contribute.
Those effective chiral NN forces are very different from the conventional
ones. But this NLO result is just an intermediate step and the final
answer has to wait for higher order contributions, which improve
systematically the observables in the 2N, 3N, ... systems at the same time.

\begin{figure}[h!]
\leftline{\mbox{\epsfysize=210mm \epsffile{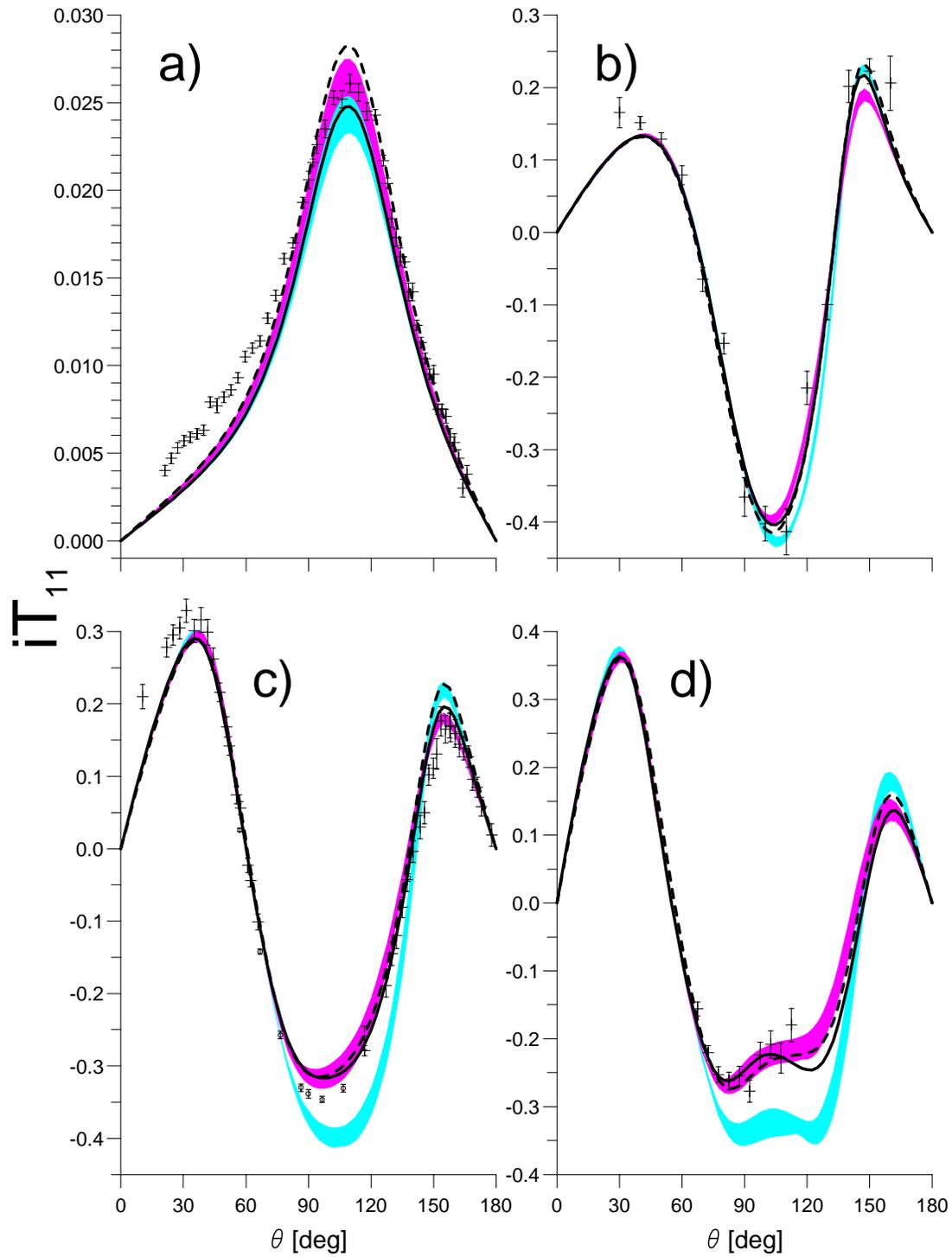}}}
\caption[ ]{The deuteron vector analyzing power $i T_{11}$ for elastic Nd scattering.
Curves and the sequence of energies as in Fig.~\protect\ref{fig1}. 
pd data at 3 MeV from~\protect\cite{ref42}, 
at 65 MeV from~\protect\cite{ref64},
at 135 MeV from~\protect\cite{ref18} (crosses), 
~\protect\cite{sakamo} (circles), 
and at 190 MeV from~\protect\cite{ref18c}.}
\label{fig3}
\end{figure}

For $i T_{11}$ shown in Fig.~\ref{fig3} the two bands are distinctly different 
and clearly the TM-band is supported by the data at the higher 
energies. In that case TM and TM' are close 
together and also TM and 
URBANA~IX except around 120~$^\circ$ at the highest energy.
The data shown in the figure for 190 MeV are taken at 197 MeV. Unfortunately 
they are absent around 120~$^\circ$.

\begin{figure}[h!]
\leftline{\mbox{\epsfysize=210mm \epsffile{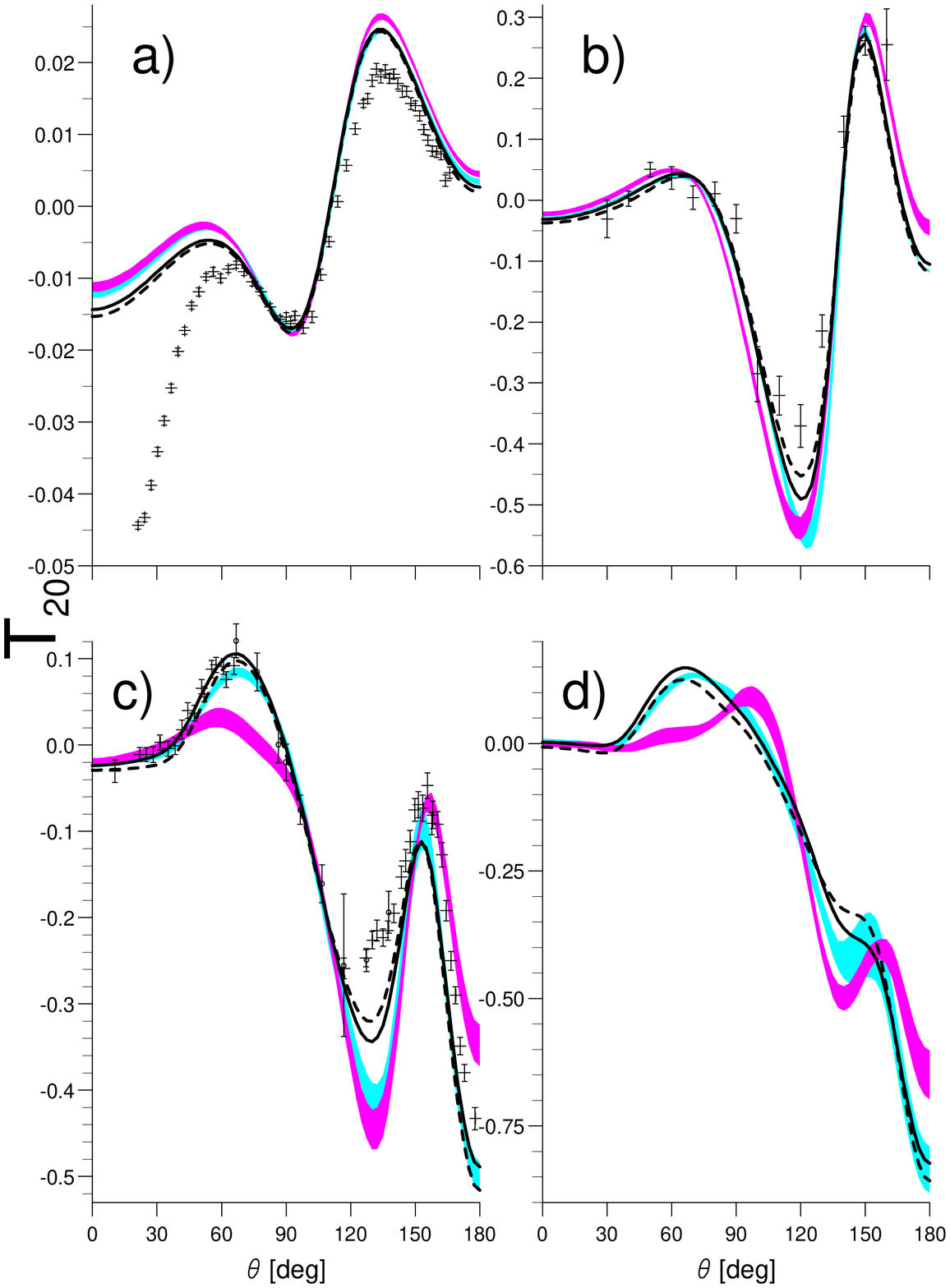}}}
\caption[ ]{The tensor analyzing power $ T_{20}$ for elastic Nd scattering.
Curves and the sequence of energies as in Fig.~\protect\ref{fig1}.
pd data  at 3 MeV from~\protect\cite{ref42},
at 65 MeV from~\protect\cite{ref64}, 
and at 135 MeV from~\protect\cite{ref18} (crosses), 
~\protect\cite{sakamo} (circles).}
\label{fig4}
\end{figure}

\begin{figure}[h!]
\leftline{\mbox{\epsfysize=210mm \epsffile{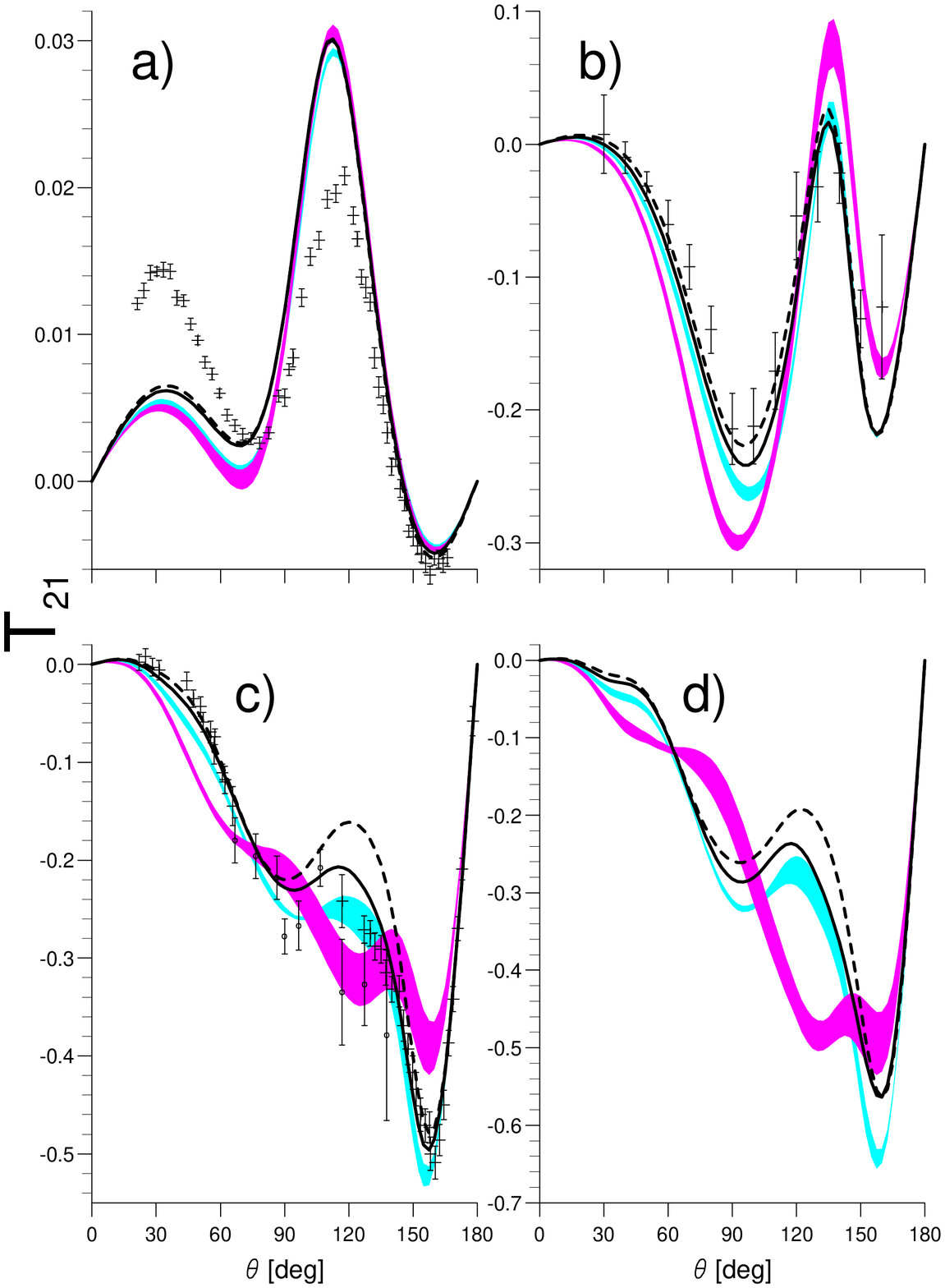}}}
\caption[ ]{The tensor analyzing power $T_{21}$ for elastic Nd scattering.
Curves and the sequence of energies as in Fig.~\protect\ref{fig1}.
pd data  at 3 MeV from~\protect\cite{ref42},
at 65 MeV from~\protect\cite{ref64},
and at 135 MeV from~\protect\cite{ref18} (crosses), 
~\protect\cite{sakamo} (circles).}
\label{fig5}
\end{figure}

\begin{figure}[h!]
\leftline{\mbox{\epsfysize=210mm \epsffile{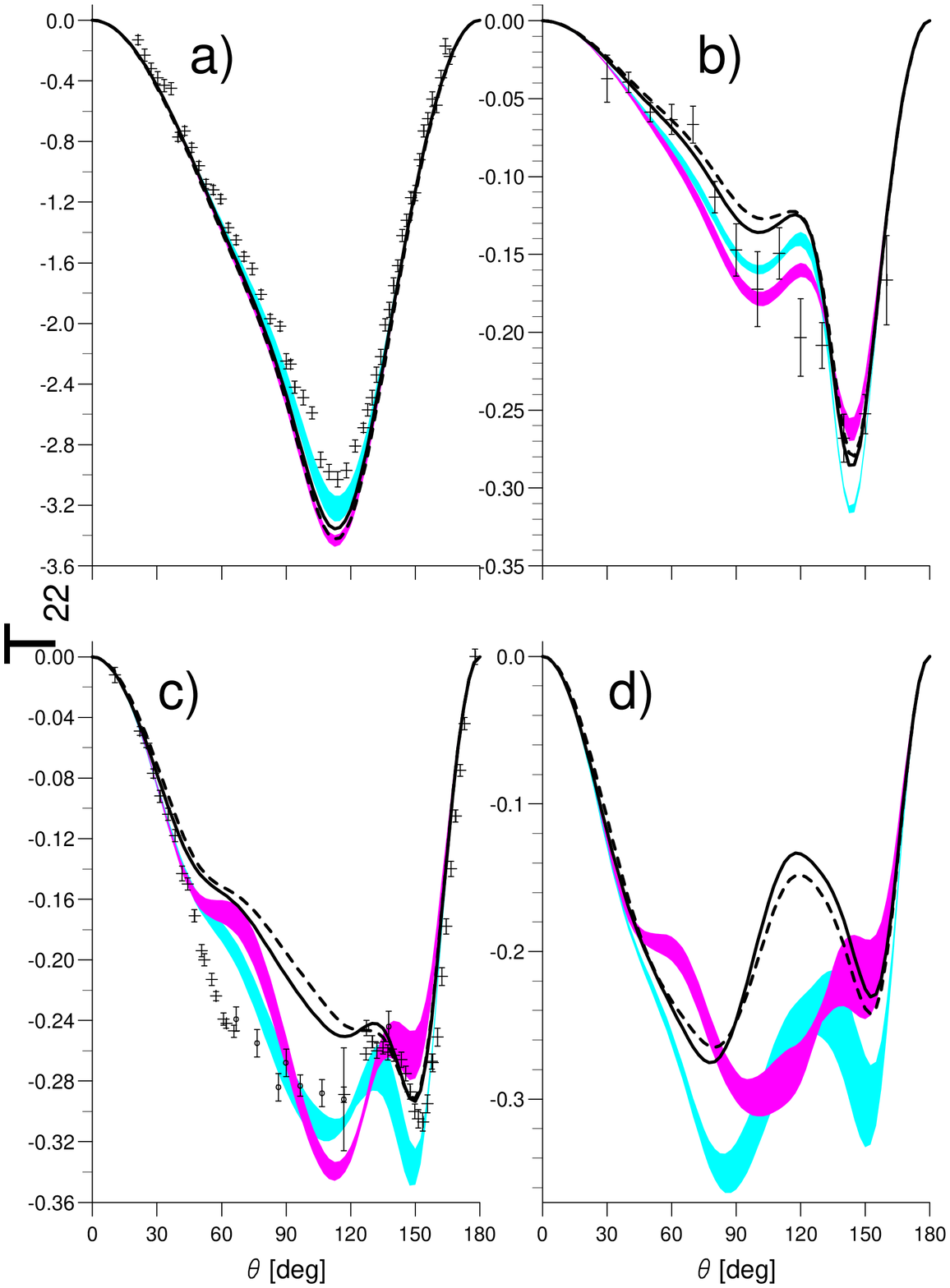}}}
\caption[ ]{The tensor analyzing power $T_{22}$ for elastic Nd scattering.
Curves and the sequence of energies as in Fig.~\protect\ref{fig1}.
pd data  at 3 MeV from~\protect\cite{ref42}, 
at 65 MeV from~\protect\cite{ref64}, 
and at 135 MeV from~\protect\cite{ref18} (crosses), 
~\protect\cite{sakamo} (circles).}
\label{fig6}
\end{figure}

Next we regard the three tensor analyzing powers $T_{20}$, $T_{21}$ 
and $T_{22}$ 
in Figs.~\ref{fig4}-\ref{fig6}. For $T_{20}$ the situation is very challenging.
At 135 MeV the data between about 120~$^\circ$-150~$^\circ$ do not agree with the overlapping
bands and above 150~$^\circ$ they lie just between the two bands. At small angles
they agree with the overlapping bands and follow then the NN force prediction.
In addition at the higher energies TM and TM' differ as well as TM 
and URBANA~IX. The strong deviations of the theory to the data at 3 MeV
is simply caused by Coulomb force effects~\cite{newKievsky}.

For $T_{21}$ the situation is different. The two bands are clearly distinct.
Again the different
3NF's predictions deviate strongly from each other. The data at 135 MeV follow more band1
than band2 and at the small angles TM' or URBANA~IX are preferred. Clearly this is a rather
contradictory situation. 

For $T_{22}$ the bands differ but the special 3NF predictions (dashed and solid 
lines) are similar 
but lie outside band2. Except at very backward and forward angles, where all curves essentially
coincide, there is disagreement with the data at $E$= 135 MeV. 
For all three $T_{2k}$'s data at 190 MeV would be very valuable,
since the various theoretical predictions differ dramatically.

\begin{figure}[h!]
\leftline{\mbox{\epsfysize=210mm \epsffile{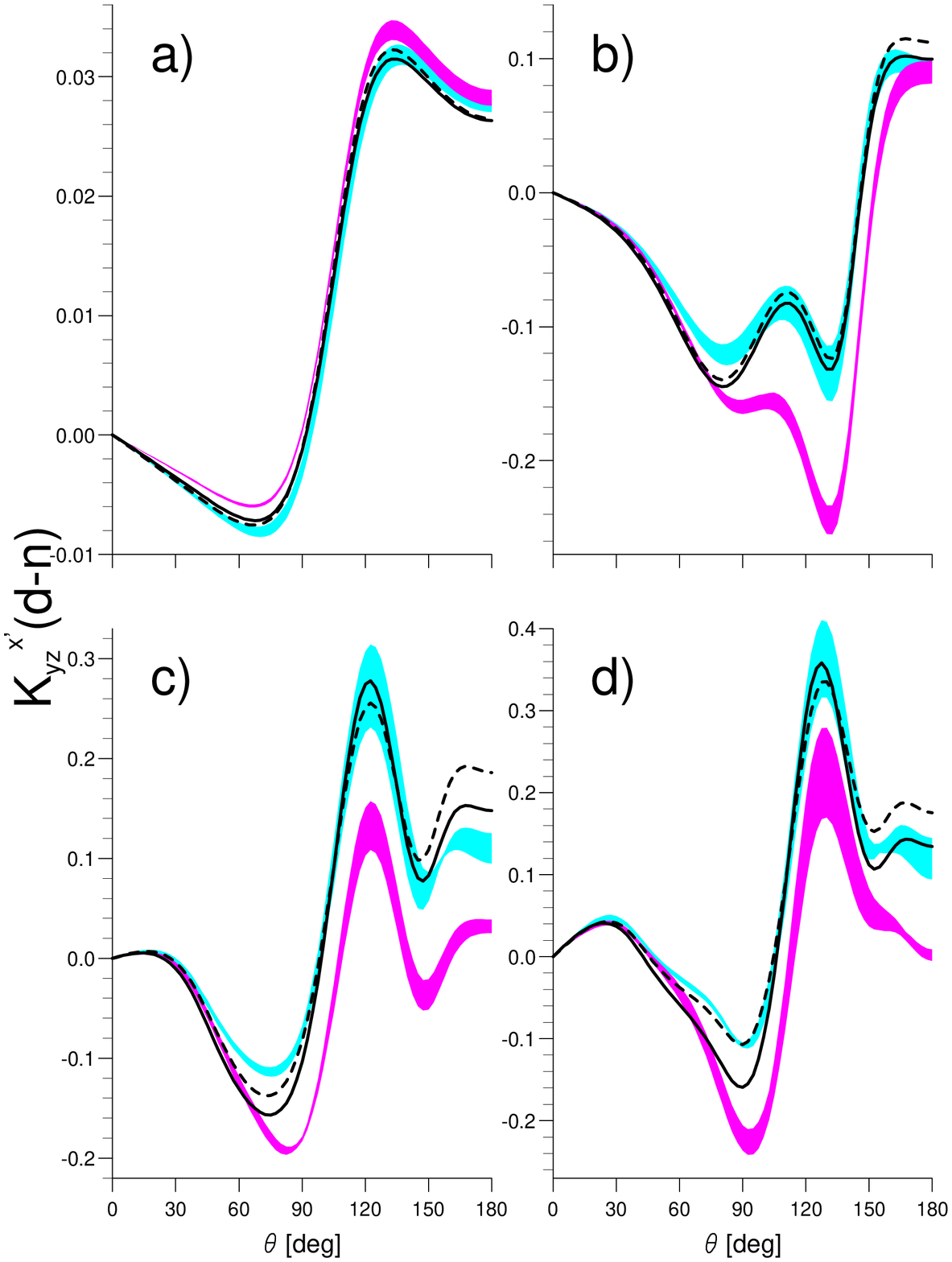}}}
\caption[ ]{The spin transfer coefficient $K_{yz}^{x '}$ 
for elastic Nd scattering.
Curves and the sequence of energies as in Fig.~\protect\ref{fig1}.}
\label{fig7}
\end{figure}

\begin{figure}[h!]
\leftline{\mbox{\epsfysize=210mm \epsffile{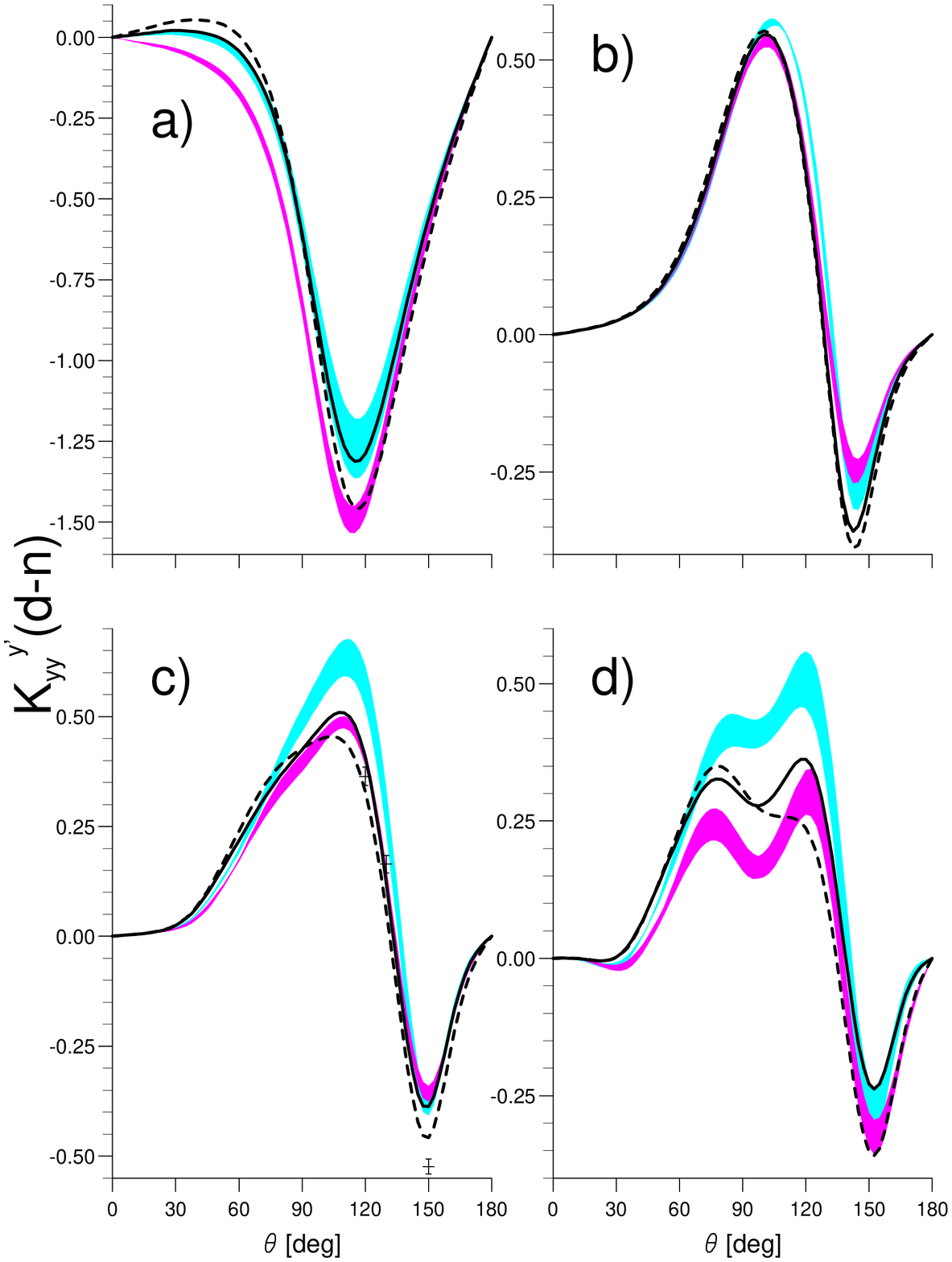}}}
\caption[ ]{The spin transfer coefficient $K_{yy}^{y '}$ for elastic Nd scattering.
Curves and the sequence of energies as in Fig.~\protect\ref{fig1}.
pd data at  135 MeV from~\protect\cite{ref18}.}
\label{fig8}
\end{figure}

\begin{figure}[h!]
\leftline{\mbox{\epsfysize=210mm \epsffile{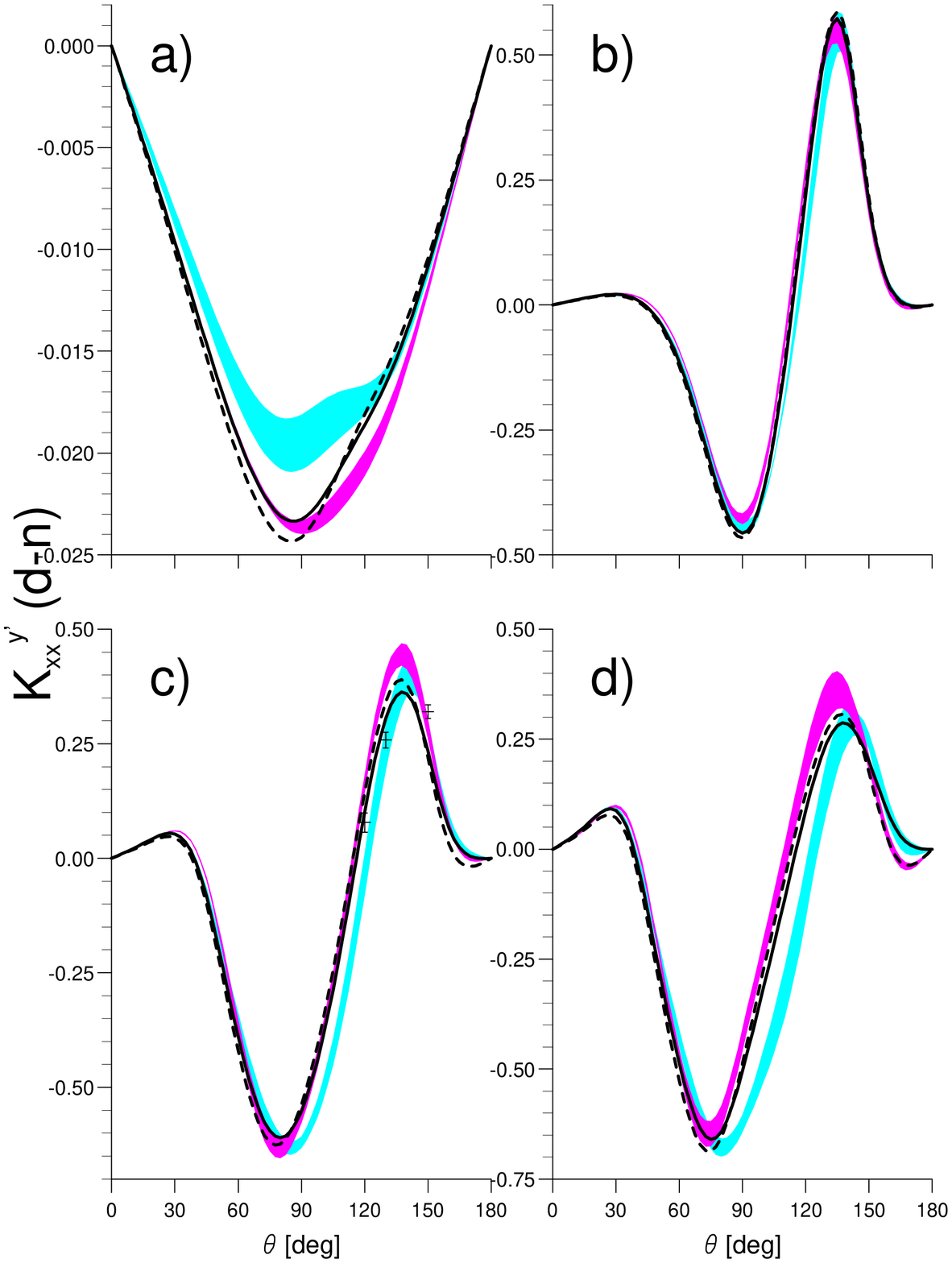}}}
\caption[ ]{The spin transfer coefficient $K_{xx}^{y '}$ for elastic Nd scattering.
Curves and the sequence of energies as in Fig.~\protect\ref{fig1}.
pd data at 135 MeV from~\protect\cite{ref18}.}
\label{fig9}
\end{figure}

\begin{figure}[h!]
\leftline{\mbox{\epsfysize=210mm \epsffile{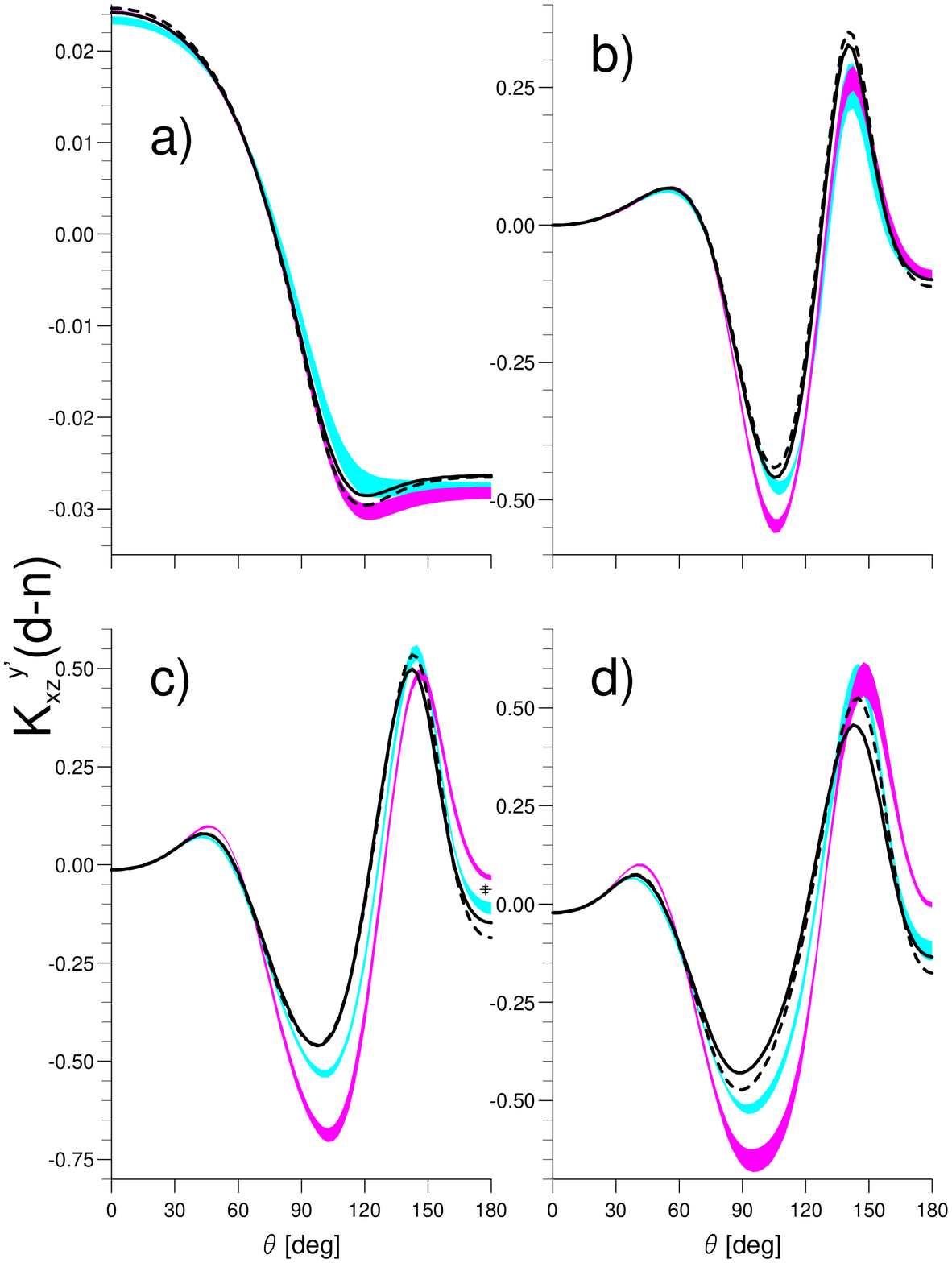}}}
\caption[ ]{The spin transfer coefficient $K_{xz}^{y '}$ for elastic Nd scattering.
Curves and the sequence of energies as in Fig.~\protect\ref{fig1}.
pd data at  135 MeV from~\protect\cite{ref18}.}
\label{fig10}
\end{figure}

\begin{figure}[h!]
\leftline{\mbox{\epsfysize=210mm \epsffile{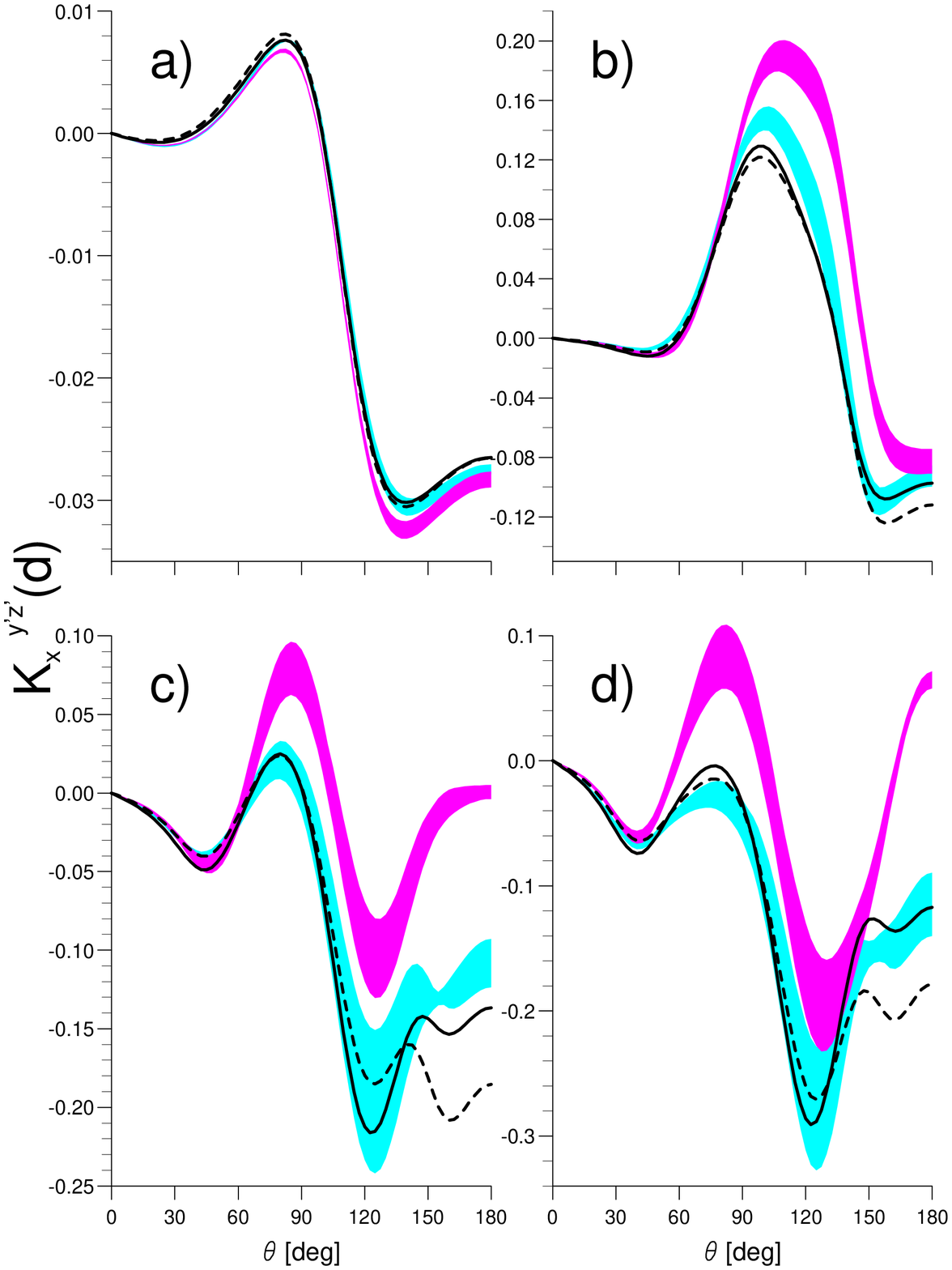}}}
\caption[ ]{The spin transfer coefficient $K_{x}^{y ' z '}$ for elastic Nd scattering.
Curves and the sequence of energies as in Fig.~\protect\ref{fig1}.}
\label{fig11}
\end{figure}

There are many spin-transfer coefficients and we selected more or less arbitrarily five of
them. In Figs.~\ref{fig7}-\ref{fig11} we show 
$K_{yz}^{x '}$,
$K_{yy}^{y '}$,
$K_{xx}^{y '}$,
$K_{xz}^{y '}$ and
$K_{x}^{y ' z '}$.
For $K_{yz}^{x '}$ the bands
strongly deviate at the higher 
energies and TM and TM' as well as TM and URBANA~IX differ
drastically.
The deviation of the bands 
from each other is less pronounced for $K_{yy}^{y '}$
and also the
different 3NF predictions are less distinct except at the highest energy.
Two data points agree with the 3NF predictions, while one, 
at 150 $^\circ$, is below any theoretical prediction.
For $K_{xx}^{y '}$ the bands differ only at the lowest energy ($E$= 3 MeV) 
significantly.  Otherwise the 3NF effects
are rather modest and also differences among the different 3NF's.
The data at 135 MeV go across the various predictions.
For $K_{xz}^{y '}$ the bands differ at the high energies 
and the two special 3NF
predictions deviate significantly from the TM predictions.
The one data point at 135 MeV lies somewhere in between.
Finally $K_{x}^{y ' z '}$ show again dramatic effects in relation to the two 
different bands. Also the special 3NF predictions are clearly 
different at the two higher energies. 

\begin{figure}[h!]
\leftline{\mbox{\epsfysize=210mm \epsffile{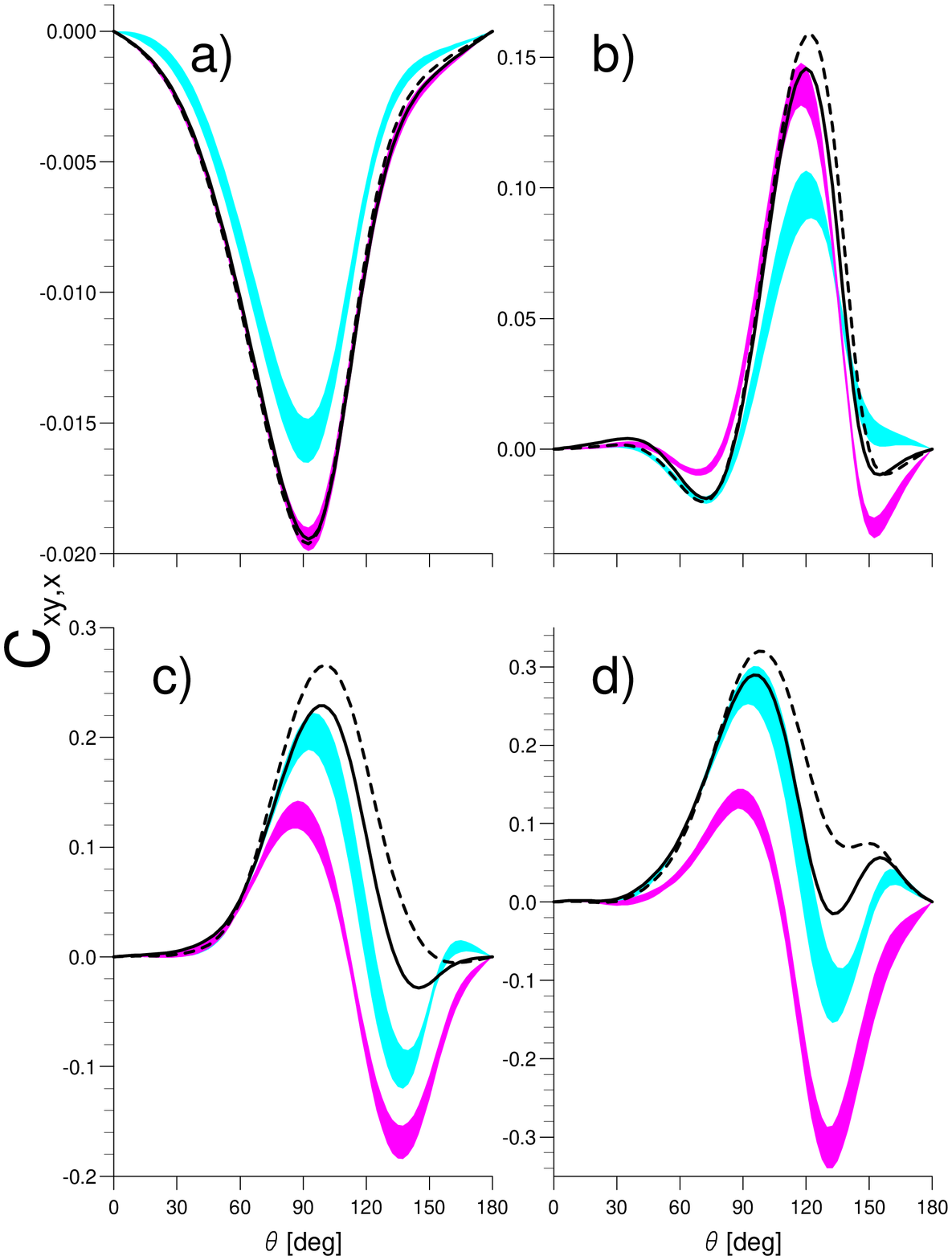}}}
\caption[ ]{The spin correlation coefficient $C_{xy, x}$ for elastic Nd scattering.
Curves and the sequence of energies as in Fig.~\protect\ref{fig1}.}
\label{fig12}
\end{figure}

\begin{figure}[h!]
\leftline{\mbox{\epsfysize=210mm \epsffile{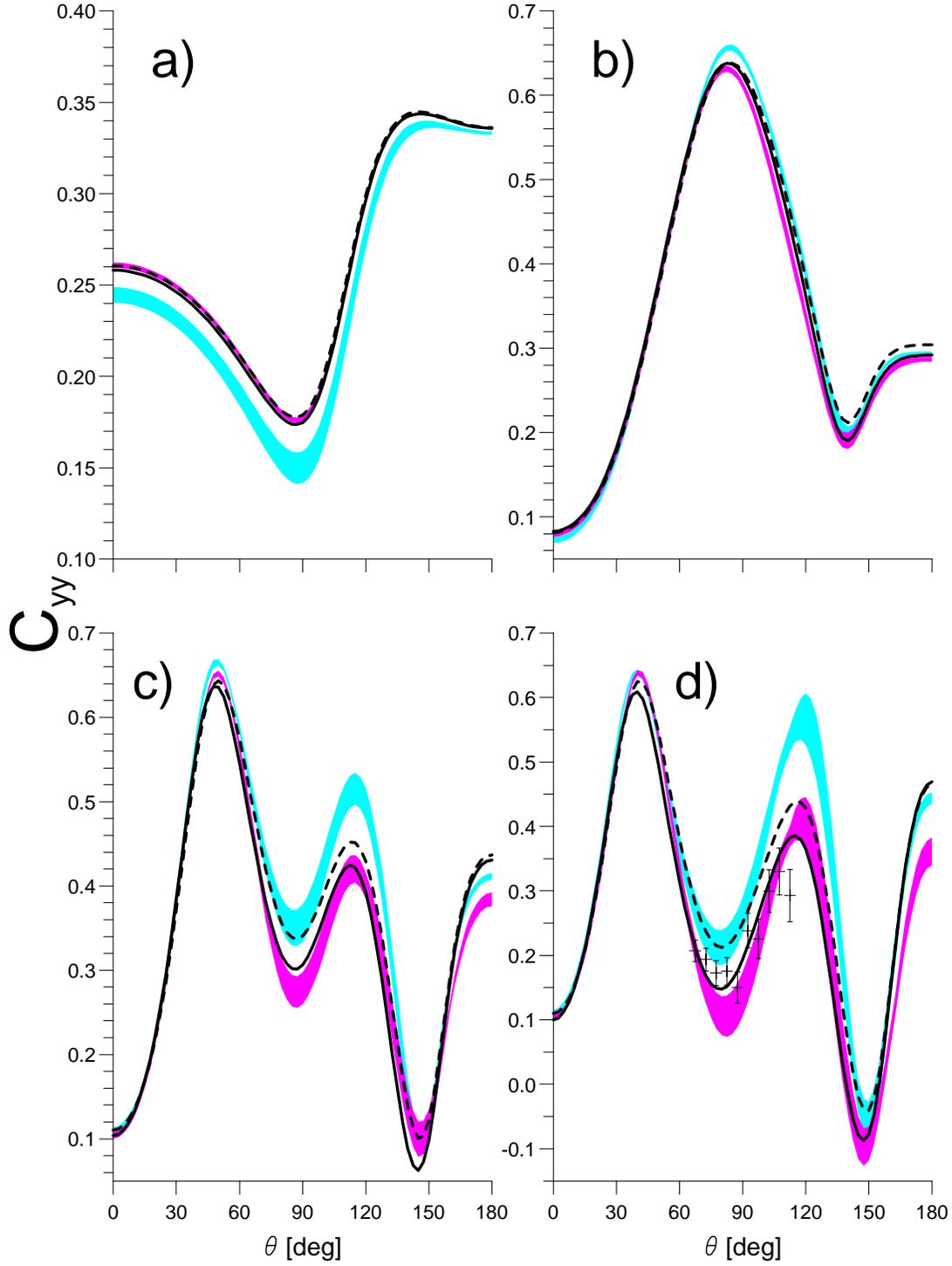}}}
\caption[ ]{The spin correlation coefficient $C_{yy}$ for elastic Nd scattering.
Curves and the sequence of energies as in Fig.~\protect\ref{fig1}.
pd data at 190 MeV from~\protect\cite{ref18c}.}
\label{fig13}
\end{figure}

\begin{figure}[h!]
\leftline{\mbox{\epsfysize=210mm \epsffile{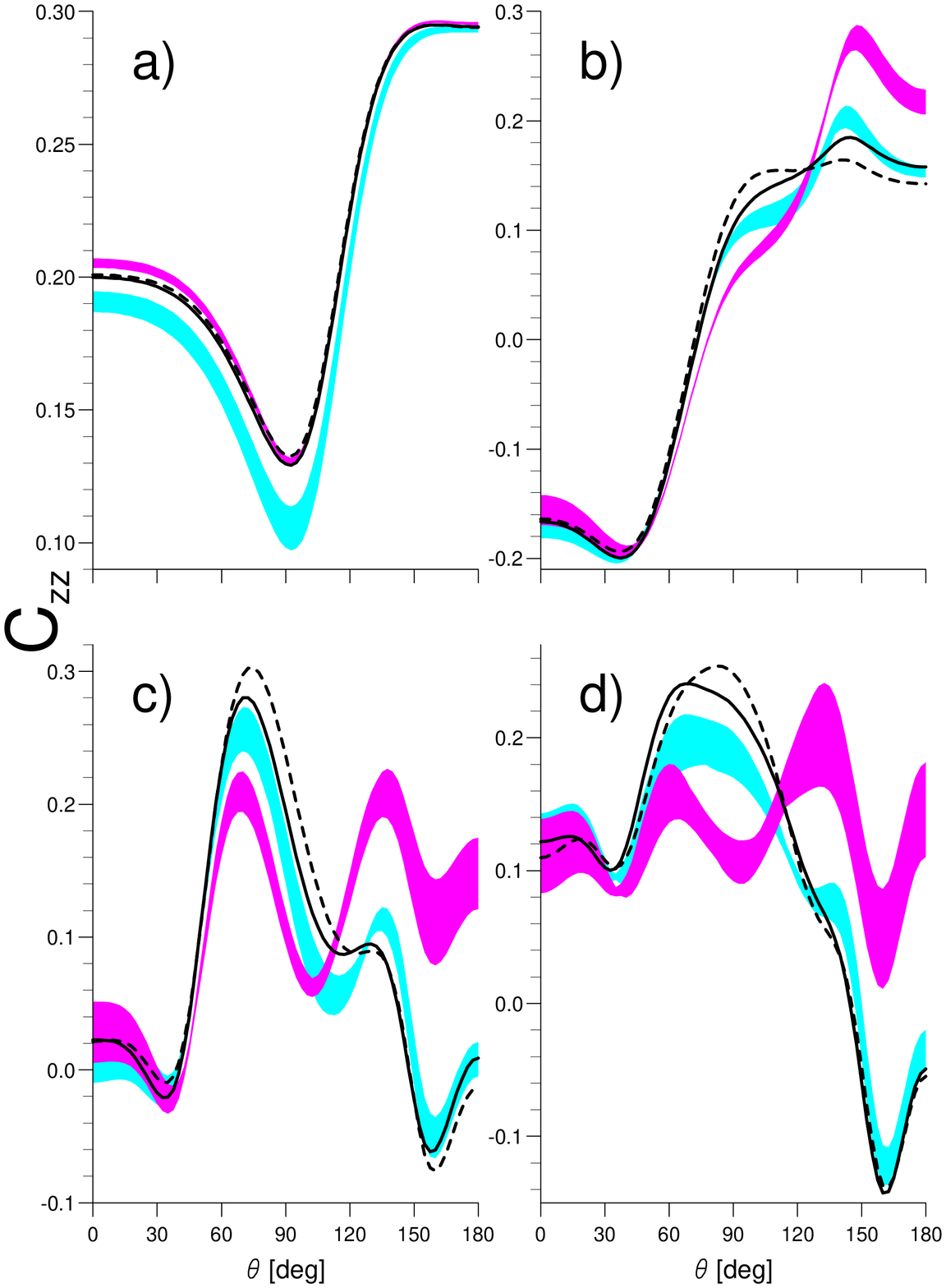}}}
\caption[ ]{The spin correlation coefficient $C_{zz}$ for elastic Nd scattering.
Curves and the sequence of energies as in Fig.~\protect\ref{fig1}.}
\label{fig14}
\end{figure}

Lastly we regard in Figs.~\ref{fig12}-\ref{fig14} three different spin correlation 
coefficients $C_{xy,x}$, $C_{yy}$ and $C_{zz}$.
The effects are dramatic for $C_{xy, x}$: the bands are quite different and also the special
3NF predictions. For $C_{yy}$ there 
are data~\cite{ref18c} at $E_p$ = 197 MeV, which we inserted 
into the figure for 190 MeV. The data  support the curves inside band2.
The AV18+URBANA~IX is 
significantly closer to experiment than CD~Bonn+TM'.
Finally for $C_{zz}$ the bands differ a lot at the high energies 
and the special 3NF predictions differ from the TM ones as well. 
One notes that URBANA~IX and TM' are similar.

\section{Summary and Conclusions}
\label{secIV}

We performed a study of popular present day 3NF models with respect 
to the effects they cause in the 3N continuum.
Based on the 
comparison of the realistic NN force predictions
alone (``band1'') to the predictions of 
all NN forces combined with the TM 3NF (``band2'') 
one sees in many spin observables very drastic effects, which should clearly be discernible 
by experiments. On top the three different 3NF models, TM, TM' and URBANA~IX combined 
(arbitrarily) with CD-Bonn and 
AV18 lead again to other very distinct predictions. 
Specifically the effects are angular and energy dependent 
becoming large at higher energies. It seems that with a sufficiently 
rich and precise data basis such diversity of effects should 
allow to nail down the proper spin structure of 3NF's.

Unfortunately there are up to now only 
few data available. The ones for the differential 
cross section support the
shift in theory caused by 3NF's. 
The existing high energy cross section data in the backward angular
region prefer the structure of the TM 3NF.
 Also the 
existing deuteron vector analyzing powers at higher energies 
support rather well the predicted 3NF effects. 
On the other 
hand this three-body interaction predicts too large effects for 
the nucleon analyzing power. This observable seems to prefer the
modified version of the Tucson-Melbourne model, TM', which is
consistent with chiral symmetry or the URBANA~IX. 
For the tensor analyzing powers
the situation is totally chaotic, for 
some scenarios one finds agreement, for others a strong
disagreement: there is no preference for any 
of them. Clearly we are at the very
beginning in investigating the spin-structure 
of the 3NF. Nevertheless the effects of
all those 3NF's are typically of the right order 
in magnitude, when they can be checked
against data but the signs are not 
yet under control. The spin transfer coefficients carry also a lot 
of information and in some of them the two bands differ very much.
Finally, the spin correlation coefficients appear also 
very informative and the very first data for $C_{yy}$
support the TM and URBANA~IX. 

Altogether the only conclusion possible is that the most popular
current 3NF models show a lot of effects and data are needed to provide
constraints.
There is hope that further theoretical work
guided by the chiral effective field theory approach will help to 
establish
the proper spin structure of the three-nucleon 
force.

\acknowledgements

This work was supported by
the Deutsche Forschungsgemeinschaft (H.K., A.N. and J.G.),
the Polish Committee for Scientific Research under Grant No. 2P03B02818 
and the Science and Technology Cooperation Germany-Poland. 
W.G. would like to thank the Foundation for Polish Science
for the financial support during his stay in Cracow.
The numerical calculations have been performed on the Cray T90 and T3E 
of the NIC in J\"ulich, Germany.

\appendix

\section{Partial wave decomposition of  the Urbana IX 3NF in momentum space.}

The Urbana 3NF in Eq.(\ref{eqURBANA3NF}) has to be put into a form suitable for the evaluation 
in partial wave decomposition. Therefore we rewrite Eq.(\ref{eqURBANA3NF}) using the 
(anti)commutator of the isospin operators
\begin{eqnarray} 
\label{eqA1}
\{ \vec{\tau}_1 \cdot \vec{\tau}_2 , \vec{\tau}_1 \cdot \vec{\tau}_3
\} & = &  2\ \vec \tau _2 \cdot \vec \tau _3  \cr
[ \vec{\tau}_1 \cdot \vec{\tau}_2 , \vec{\tau}_1 \cdot \vec{\tau}_3 ]
& = &  2i \ \vec \tau_1 \cdot \vec \tau _2 \times \vec \tau _3 
\end{eqnarray}
$V_4^{(1)}$ reads then
\begin{eqnarray}
\label{eqA2}
V_4^{(1)} & = & A_{2\pi} 
           \ 2 \left( \vec {\tau_2} \cdot \vec {\tau_3} 
                       + { i \over 4 } \  \vec {\tau_1}\cdot \vec {\tau_2} \times \vec {\tau_3} 
                         \right) \ X_{12} X_{13} \cr
          & & + A_{2\pi} 
           \ 2 \left( \vec {\tau_2} \cdot \vec {\tau_3} 
                       - { i \over 4 } \  \vec {\tau_1}\cdot \vec {\tau_2} \times \vec {\tau_3} 
                         \right) \ X_{13} X_{12} \cr
          & & + U_{0} \ T^2_{\pi} (r_{12}) \ T^2_{\pi} (r_{13}) \cr
          & = & A_{2\pi} 
           \ 2 \left( \vec {\tau_2} \cdot \vec {\tau_3} 
                       + { i \over 4 } \  \vec {\tau_1}\cdot \vec {\tau_2} \times \vec {\tau_3} 
                         \right) \ X_{12} X_{13} \cr
          & & + A_{2\pi} 
           \ 2 \left( \vec {\tau_2} \cdot \vec {\tau_3} 
                       - { i \over 4 } \  \vec {\tau_1}\cdot \vec {\tau_2} \times \vec {\tau_3} 
                         \right) \ X_{13} X_{12} \cr
          & & + { U_{0} \over 2 }  \ \left( 
                   T^2_{\pi} (r_{12}) \ T^2_{\pi} (r_{13}) 
                  +T^2_{\pi} (r_{13}) \ T^2_{\pi} (r_{12}) \right) 
\end{eqnarray}
Because the $T^2$ operators commute which each other, we could choose the symmetrical form 
in the last line.

The aim is to find matrix elements with respect to our standard basis states \cite{gloecklebook}
\begin{equation}
\label{eqA3}  
| p \ q \alpha > _i \equiv | p \ q \ (ls)j \ (\lambda { 1 \over 2 }) J \ (jJ) {\cal J} {\cal M} 
                                     \ \ (t  { 1 \over 2 } ) T M_T > _i 
\end{equation}
Here the magnitudes of the Jacobi momenta of the subsystem  and the
outer particle
are $p$ and $q$, respectively. The angular dependence is expanded in
partial waves. Corresponding to $p$ and $q$ we introduce angular
momenta $l$ and $\lambda$. As indicated in Eq.(\ref{eqA3}) the orbital 
angular momenta couple with the spin of the subsystem $s$ and the
outer particle ${1\over 2}$ to the total spin of the subsystem $j$ and 
outer particle $J$. These angular momenta are combined to the total
spin $\cal J$ and its third component $\cal M$. The total isospin of the
subsystem $t$ couples with the isospin of the outer particle ${1\over 2}$
 to $T$ and
its third component $M_T$. Because we will make use of several sets
of Jacobi momenta, we append an index $i$ which gives the number of
the outer particle.  

According to Eqs.(\ref{eqT}) and (\ref{eqU}) the operator $V_4^{(1)}$ 
acts on  completely antisymmetric 3N states 
$ | \chi > = (1+P) \ | \phi >$ or $ | \chi > = (1+P) \ G_0 T $. 
In the next paragraphs we would like to establish some consequences of this fact. 

To that aim we introduce a short hand notation for our basis states:
\begin{equation}
\label{eqA4}
| (jk)i >  \equiv | p \ q \alpha >_i 
\end{equation}
In contrast to Eq.(\ref{eqA3}) this definition fixes the ordering
within the subsystem. $(ij)$ is understood to fix the momentum vector
to  $\vec p = {1\over2} ( \vec k_i-\vec k_j)$ and the spin and isospin 
coupling within the subsystem to $(s_i s_j) s$ and $(t_i t_j)t$. In
consequence we can distinguish $|(ij)k>$ and $|(ji)k>$ in this
notation. In the partial wave decomposition there is a simple phase
relation connecting both states 
\begin{equation} 
\label{eqA5}
 |(ij)k> = (-) ^ { l+s+t} \ |(ji)k> \equiv (-)^{(ij)} \ |(ji)k> 
\end{equation}
Of course all sets of basis states are complete, therefore we can expand 
the incoming state in several ways
\begin{eqnarray}
\label{eqA6} 
| \chi > & = & \intsum   |(31)2 > <(31)2 | \chi > \cr
         & = & \intsum   |(12)3 > <(12)3 | \chi > 
\end{eqnarray}
Due to the  total antisymmetry of $| \chi >$, its matrix elements
are equal 
\begin{equation} 
<(31)2 | \chi >=<(12)3 | \chi >=<(23)1 | \chi >
\end{equation}
In consequence 
we can expand the different terms in Eq.(\ref{eqA2}) on the right hand side in 
different Jacobi coordinates. Let us begin with the two $U_0$ 
terms in Eq.(\ref{eqA2}):
\begin{eqnarray}
\label{eqA7}
M_3 & = &    \phantom{+} { 1 \over 2 } \ <(23)1 | (12)3 ''> \ < (12)3 ''| T^2_{\pi} (r_{12})  | (12)3 '''> 
\cr 
& & < (12)3 '''| (31) 2 ''''> \ 
         < (31) 2''''| T^2_{\pi} (r_{13}) | (31) 2 '>   \cr
& & + { 1 \over 2 } <(23)1 | (31)2 ''> \ < (31)2 ''| T^2_{\pi} (r_{13})  | (31)2 '''> 
       \cr 
& &     < (31)2 '''| (12) 3 ''''> \
         < (12) 3''''| T^2_{\pi} (r_{12}) | (12) 3 '>   
\end{eqnarray}
In this equation we introduced several  completeness relations and
projected on two kinds of incoming states keeping in mind that one can add 
up the matrix elements because 
of the total antisymmetry of $|\chi>$. In the way we inserted the 
completeness relations in Eq.(\ref{eqA7}), the matrix elements of
$T^2_{\pi}$ are evaluated in their natural coordinates. In this form the operator is
diagonal in the quantum numbers and momenta of the outer particle and
additionally it conserves the symmetry with respect to the interchange 
of the particles of the subsystem.

The matrix element 
$< (12) 3| T^2_{\pi} (r_{12}) | (12) 3 '>$ 
depends on Jacobi coordinates and quantum numbers which single out the 
subsystem (12). By renumbering the particles one finds  
\begin{equation} 
< (12) 3| T^2_{\pi} (r_{12}) | (12) 3 '>= < (31) 2| T^2_{\pi}
(r_{13}) | (31) 2 '>
\end{equation}
Note that the matrix element on the right hand side
depends on coordinates which differ from the ones on the left hand side. Though 
the matrix elements are equal, one has to keep in mind that the
actual meaning of the momenta and quantum numbers is different on both sides.  

In the same manner one can find an important
relation between the cyclic and anti cyclic  transformations 
\begin{eqnarray} 
\label{eqA8}
<(ij)k| (jk) i '> = <(ik)j | (kj) i '> = (-)^{(ik)}(-)^{(kj)'} <(ki)j| (jk) i '> 
\end{eqnarray}
We would like to emphasize that the transformation itself doesn't conserve the symmetry
of the subsystem and therefore the completeness relation in $''$ and
$'''$ states in Eq.(\ref{eqA7}) have to include also the unphysical
symmetric states in the two-body subsystems. 
With the help of Eq.(\ref{eqA8}) and renumbering the particles in the second
part of Eq.(\ref{eqA7}) one finds 
\begin{eqnarray}
\label{eqA9}
M_3 & = &    \phantom{+} { 1 \over 2 } \ <(23)1 | (12)3 ''> \ < (12)3 ''| T^2_{\pi} (r_{12})  | (12)3 '''> 
\cr 
& & < (12)3 '''| (31) 2 ''''> \ 
         < (31) 2''''| T^2_{\pi} (r_{13}) | (31) 2 '>   \cr
& & + { 1 \over 2 } (-)^{(23)}(-)^{(12)''} (-)^{(12)'''}(-)^{(31)''''} <(23)1 | (12)3 ''> 
             \ < (12)3 ''| T^2_{\pi} (r_{12})  | (12)3 '''> 
       \cr 
& &     < (12)3 '''| (31) 2 ''''> \
         < (31) 2''''| T^2_{\pi} (r_{13}) | (31) 2 '>   
\end{eqnarray}
Because $T^2_{\pi}$ conserves the symmetry of the subsystem
$(-)^{(12)''} =(-)^{(12)'''}$ and $(-)^{(31)''''}
=(-)^{(31)'}$. Therefore Eq.(\ref{eqA9}) reduces to 
\begin{eqnarray}
\label{eqA10}
M_3 & = &  { 1 \over 2 } \
\left(1+(-)^{(23)}(-)^{(31)'}\right) 
       \ <(23)1 | (12)3 ''> \ < (12)3 ''| T^2_{\pi} (r_{12})  | (12)3 '''> 
\cr 
& & < (12)3 '''| (31) 2 ''''> \ 
         < (31) 2''''| T^2_{\pi} (r_{13}) | (31) 2 '>   
\end{eqnarray}
The incoming state is antisymmetric in the subsystem (31) hence $(-)^{(31)'}=-1$. Therefore
$M_3$ is zero for outgoing states which are symmetric in the subsystem (23) and equals twice the first
part for the antisymmetric outgoing states. We restrict the outgoing
states to antisymmetric ones. Then it is justified to write
\begin{eqnarray}
\label{eqA11}
M_3 & = &    
   <(23)1 | (12)3 ''> \ < (12)3 ''| T^2_{\pi} (r_{12})  | (12)3 '''> 
\cr 
& & < (12)3 '''| (31) 2 ''''> \ 
         < (31) 2''''| T^2_{\pi} (r_{13}) | (31) 2 '>   
\end{eqnarray}
We would like emphasize again that we restrict the outgoing, incoming
and $''''$ states to physical, antisymmetric states in the subsystems. Due to the
coordinate transformations, the $''$ and $'''$ sums are not restricted
anymore to antisymmetric subsystem states and the completeness
relations run over all symmetries. 

Lets turn to the $A_{2\pi}$ parts now. In the same manner we define
the matrix elements of the $A_{2\pi}$ parts
\begin{eqnarray} 
\tilde M_3 & = &    \phantom{+} \ <(23)1 | (12)3 ''> \ < (12)3 ''| X_{12}  | (12)3 '''> 
\cr 
& & < (12)3 '''| 2 \left( \vec {\tau_2} \cdot \vec {\tau_3} 
                       + { i \over 4 } \  \vec {\tau_1}\cdot \vec {\tau_2} \times \vec {\tau_3} 
                         \right) | (31) 2 ''''> \ 
         < (31) 2''''| X_{13} | (31) 2 '>   \cr
& & + <(23)1 | (31)2 ''> \ < (31)2 ''| X_{13} | (31)2 '''> 
       \cr 
& &     < (31)2 '''| 2 \left( \vec {\tau_2} \cdot \vec {\tau_3} 
                       - { i \over 4 } \  \vec {\tau_1}\cdot \vec {\tau_2} \times \vec {\tau_3} 
                         \right) |(12) 3 ''''> \
         < (12) 3''''| X_{12} | (12) 3 '>   
\end{eqnarray}
The isospin operators commute with the $X_{ij}$ operators. We combined 
them with the inner transformation matrix. Because $ \vec {\tau_2}
\cdot \vec {\tau_3} $  is symmetric with respect to the interchange of 
particle 2 and 3 and $\vec {\tau_1}\cdot \vec {\tau_2} \times \vec
{\tau_3}$ is antisymmetric, the steps leading to Eq.(\ref{eqA8}) can
be repeated with the isospin matrix element. In addition there is a
sign change for the antisymmetric part of the isospin operators.
\begin{eqnarray}  
& &  < (12)3 '''| 2 \left( \vec {\tau_2} \cdot \vec {\tau_3} 
                       + { i \over 4 } \  \vec {\tau_1}\cdot \vec {\tau_2} \times \vec {\tau_3} 
                         \right) | (31) 2 ''''> \cr 
& & \quad = (-)^{(31)'''} (-)^{(12)''''}  < (31)2 '''| 2 \left( \vec {\tau_2} \cdot \vec {\tau_3} 
                       - { i \over 4 } \  \vec {\tau_1}\cdot \vec {\tau_2} \times \vec {\tau_3} 
                         \right) | (12) 3 ''''>   
\end{eqnarray}
The symmetry of the $X_{ij}$'s leads to equal phases on both sides
of their matrix elements: $(-)^{(31)''''}=(-)^{(31)'}$ and
$(-)^{(12)''}=(-)^{(12)'''}$.
The matrix element of the $A_{2\pi}$ parts reduces to 
\begin{eqnarray} 
\tilde M_3 & = & \left(1 + (-)^{(23)} (-)^{(31)'} \right) \ 
      <(23)1 | (12)3 ''> \ < (12)3 ''| X_{12}  | (12)3 '''> 
\cr 
& & < (12)3 '''| 2 \left( \vec {\tau_2} \cdot \vec {\tau_3} 
                       + { i \over 4 } \  \vec {\tau_1}\cdot \vec {\tau_2} \times \vec {\tau_3} 
                         \right) | (31) 2 ''''> \ 
         < (31) 2''''| X_{13} | (31) 2 '>   
\end{eqnarray}
and its is again justified to restrict the outgoing
states to antisymmetric subsystems and write  
\begin{eqnarray} 
\tilde M_3 & = & 2 \ 
      <(23)1 | (12)3 ''> \ < (12)3 ''| X_{12}  | (12)3 '''> 
\cr 
& & < (12)3 '''| 2 \left( \vec {\tau_2} \cdot \vec {\tau_3} 
                       + { i \over 4 } \  \vec {\tau_1}\cdot \vec {\tau_2} \times \vec {\tau_3} 
                         \right) | (31) 2 ''''> \ 
         < (31) 2''''| X_{13} | (31) 2 '>   
\end{eqnarray}

It is standard to work out the two-body partial wave matrix elements for $X_{ij}$  and $T^2_{\pi}$ 
in terms of Bessel transforms. Also the isospin and coordinate 
transformation matrix elements are standard 
and we refer to \cite{ref37}.


\begin{table}[htb]
\label{TAB1}
\begin{tabular}{@{}ll}
\hline
           & $\Lambda \ [m_\pi] $ \\
\hline
 CD Bonn+TM   & 4.856  \\
 AV18+TM      & 5.215  \\
 Nijm I+TM    & 5.120  \\
 Nijm II+TM   & 5.072  \\
 Nijm'93+TM   & 5.212  \\
\hline
\end{tabular}
\caption[]{The cut-off parameters $\Lambda$ from Eq.~(\protect\ref{eqH}) used 
in the given potential combinations.}
\end{table}


\end{document}